# Alpha Cygni Variables as Seen from the Transiting Exoplanet Survey Satellite

*Joyce A. Guzik[1], Claire Whitley[1,2], Nova Moore[1,3], Madeline Marshall[1], and Jason Jackiewicz[4]*

[1]Los Alamos National Laboratory, Los Alamos, NM (joy@lanl.gov); [2]University of New Mexico, Albuquerque, NM; [3]New Mexico Institute of Mining and Technology, Socorro, NM; [4]New Mexico State University, Las Cruces, NM

**Subject Keywords**

Alpha Cygni Variables; AAVSO International Database; NASA TESS Spacecraft

**Abstract**

The Alpha Cygni (ACYG) variables are blue-white supergiants which display low-amplitude brightness variations of around 0.1 magnitude. The prototype Deneb shows quasi-periodic variations of around 12 days, interrupted by intervals of erratic variability, and occasionally large excursions in amplitude. To gain insight on the behavior of these variables, we examined 27-day light curves from the Transiting Exoplanet Survey Satellite (TESS) for 75 ACYG variables south of the ecliptic plane which are being revisited by TESS in 2025-2026. We use the web-based TESS Extractor app for screening TESS light curves. We identified ten stars with similarities to Deneb that may be good candidates for ground-based monitoring. We approximated the location of these stars on the Hertzsprung-Russell diagram, and find most lie below the Luminous Blue Variables, are cooler than the β Cephei variables, and are hotter than the RV Tauri stars. We also compare light curves processed with several different pipelines available on the Mikulski Archive for Space Telescopes (MAST) and comment on their utility for ACYG stars.

## 1. Introduction

Alpha Cygni (ACYG) variables are luminous supergiants of spectral types O, B, and A which show low-amplitude (around 0.1 mag or less) photometric variability (e.g., van Genderen et al. 1989, Richardson et al. 2010). The prototype α Cyg (Deneb), with visual magnitude 1.21-1.29, exhibits alternating phases of quasi-periodic and erratic variability (see, e.g., Abt et al. 2023, Guzik et al. 2024a, 2024b, 2025), raising the question of whether similar behaviors are common among other members of the class. These variations have potential to provide insights into the evolutionary state and processes in the envelopes and atmospheres of massive stars nearing the ends of their lives.

The small-amplitude variations and quasi-periodicities of days to more than a month make ACYG variables challenging targets for ground-based observing. Since the variations do not strictly repeat in period or amplitude like classical variable stars such as Cepheids or RR Lyrae stars, it is not possible to phase data from sparse observations to create a light curve. Light curves spanning several months or years are necessary to search for abrupt excursions or changes from erratic to quasi-periodic variability, as seen in Deneb. Ideally, a photometric data point around once per night is required, but seasonal visibility and cloudy skies make this goal difficult to attain. In addition, visual observing and perhaps even CCD photometry cannot obtain the necessary precision (0.01 mag) to characterize ACYG variations. The AAVSO Photoelectric Photometry (PEP) team has had good success in observing Deneb and will be targeting other ACYG variables (see Calderwood and van Ballegoij, these proceedings).

Ongoing observations by the NASA Transiting Exoplanet Surve Satellite (TESS, Ricker et al. 2015) spacecraft may be useful for characterizing ACYG variables. TESS was launched April 18, 2018 into a 13.7-day lunar resonance orbit around Earth. TESS's four CCD cameras have 2048 x 2048 pixels, with 21 arc seconds per pixel, and together cover 24° x 96° strips of the sky, called sectors. Photometric time series are taken for each sector for 27.4 days at a time. TESS's primary mission is to search for planets around nearby, mostly M-dwarf, stars using the planet transit method. 705 confirmed planets and 7703 candidates have been discovered as of October 2025.[1]

TESS surveyed sectors south and north of the ecliptic plane during its first two years of operation. The cadence for full-frame images was 30 minutes for years 1-2, then was shortened to 10 minutes for years 3-4, and subsequently has been 3.3 minutes. Observations through Sector 90, taken in March/April 2025, were available by the end of August when this paper was being prepared.

To obtain TESS light curves, the raw pixel data must be processed to subtract backgrounds, check for contamination from nearby stars in the field of view, detrend, remove outliers, and calculate error bars. We used the web-based TESS Extractor app (Serna et al. 2021) to screen TESS light curves and identify stars with behavior similar to Deneb's. The Mikulski Archive for Space Telescopes (MAST) includes light-curve products processed using different processing pipelines. For ACYG variables, it is important to remove artificial trends while not removing real variability. We compare example light curves from several pipelines along with the TESS Extractor results and discuss our preferences for obtaining TESS ACYG light curves.

**2. TESS Observations of ACYG Variables**

At prior AAVSO meetings, we reported TESS observations of Deneb. Abt et al. (2023), Guzik et al. 2024a,b and Guzik et al. 2025 discuss the special 2-minute cadence high-level science product (HLSP) light curves available at the Mikulski Archive for Space Telescopes (MAST).[2] The Deneb light curve shows mostly irregular variations in Sector 41, and resumption of the quasi-periodic 12-day variations during Sector 55, continuing into Sector 56 (Figure 1).

---

[1] https://exoplanetarchive.ipac.caltech.edu/
[2] https://archive.stsci.edu

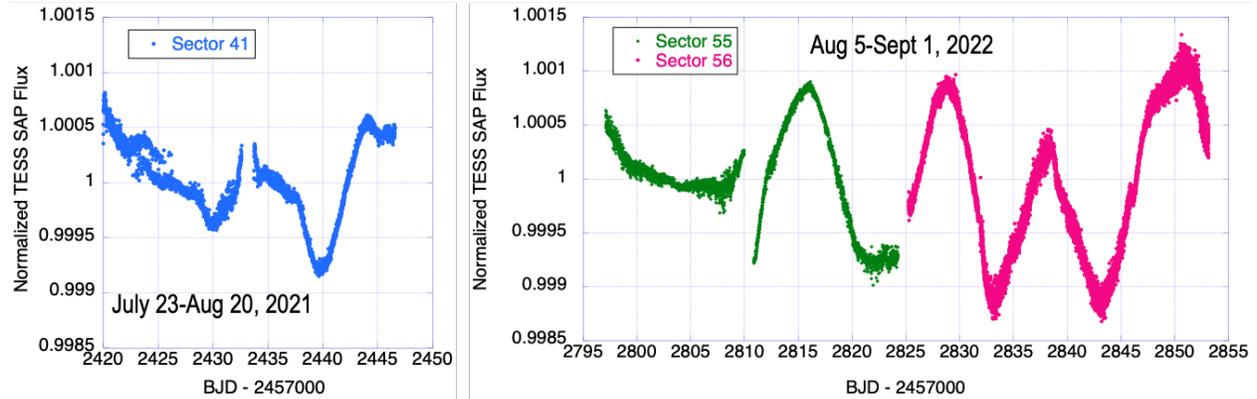

*Figure 1. Deneb photometry from TESS spacecraft as presented by Abt et al. (2023). Variations with quasi-period around 12 days resume around day 2810 in Sector 55 data.*

In March 2025, we proposed via the TESS General Investigator Program to obtain light curves for 75 out of 186 stars classified as ACYG variables in the AAVSO International Variable Star Index[3] which will be observable by TESS during Cycle 8, Sectors 97-107, Sept. 2025-Sept. 2026. These targets, positioned south of the ecliptic plane, span TESS magnitudes 0.7 (Rigel) to 12.6. Figure 2 shows the planned observing fields in ecliptic and equatorial coordinate views. For this Cycle, the fields of view of consecutive sectors will overlap at middle latitudes, so it will be possible to obtain coverage for some ACYG variables for at least two consecutive sectors, or 54 days, an advantage for monitoring the changing behavior of these stars.

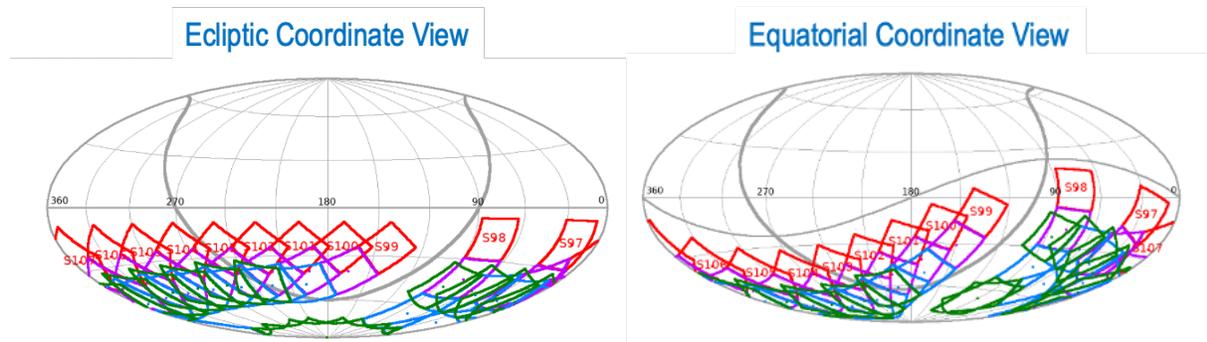

*Figure 2. TESS fields of view for Cycle 8, Sectors 97-107, to be observed south of the ecliptic plane, September 2025-September 2026.[4]*

**3. Light Curve Analysis Using TESS Extractor**

All 75 targets have one or more prior 27-day sector of TESS observations. We used the web-based application TESS Extractor[5] (Serna et al. 2021; Brasseur et al. 2019) to screen these data.

---

[3] https://vsx.aavso.org/
[4] https://tess.mit.edu/observations/
[5] https://www.tessextractor.app/

The TESS Extractor graphical interface is easy to use, and one does not need to know programming languages such as python, or how to process target pixel files or full-frame images. A disadvantage is that obtaining a light curve for a single 27-day sector for a given star via the interface may take several minutes. TESS Extractor has a bulk download request option, but this option was unavailable during the summer of 2025 when we were working on this project.

Figure 3 shows a screen shot of the TESS Extractor application page. While the TESS full-frame images are partially corrected before release for, e.g., CCD camera effects and cosmic rays, additional detrending is needed to remove additional systematic effects, mainly scattered light from the Earth and Moon, but also spacecraft pointing jitter and changes in detector sensitivity because of temperature variations (see, e.g., Hattori et al. 2022). To obtain detrended data, one should select the option noted by the red arrow. TESS Extractor uses a detrending approach inspired by PyKe (Still and Barklay 2012), a legacy module designed for *Kepler* spacecraft data designed for reduction and analysis of Simple Aperture Photometry data. Target names can be entered by TESS Input Catalog (TIC) number, but the software also recognizes many other common star names, such as Deneb or Rigel. After entering the star name, one then clicks the 'Resolve Target' button, and, after the target is found, one can choose using the drop-down menu which TESS sector to process.

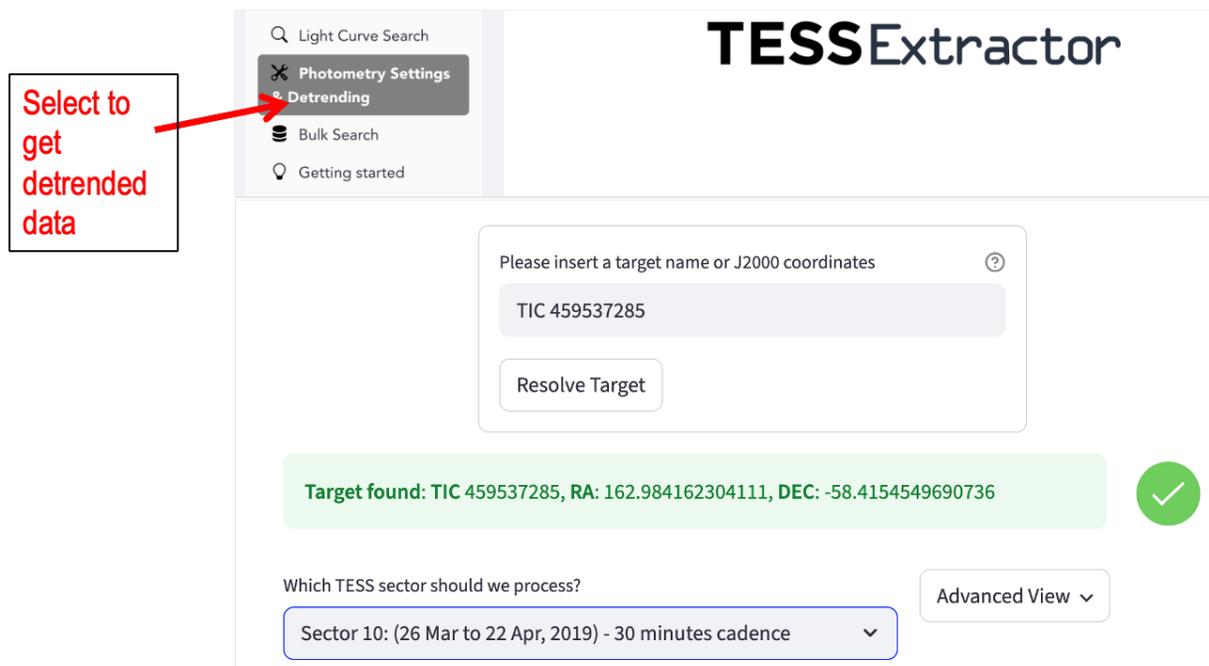

*Figure 3. Screen shot of user interface for TESS Extractor[6] online interactive application.*

Figure 4 shows a screen shot of the resulting 'simple aperture photometry' light curve (blue dots) and the trends based on photometry of other pixels in the field of view surrounding the star. Scrolling further in the analysis screen, the detrended light curve, a Lomb-Scargle periodogram, and some plots phased by the significant periods are shown (Figure 5). Text files with the time-series detrended light curve and the Lomb-Scargle periodogram can be downloaded using the buttons below the plots.

---

[6] https://www.tessextractor.app/

Note that the periods in Figure 5 (right) are labeled as 'rotation' periods. However, for ACYG variables, these are not rotation periods, just significant periods in the 27-day light curve. For TIC 459537285, the light curve is not strictly periodic, although periods of 4.9 and 7.3 days are identified, but a time-series length of only 27 days is insufficient to characterize the variability of this star. The period identified at 0.010 days may be an alias of the 30-minute cadence, since 30 minutes = 0.0208 days, or two times 0.0104 days.

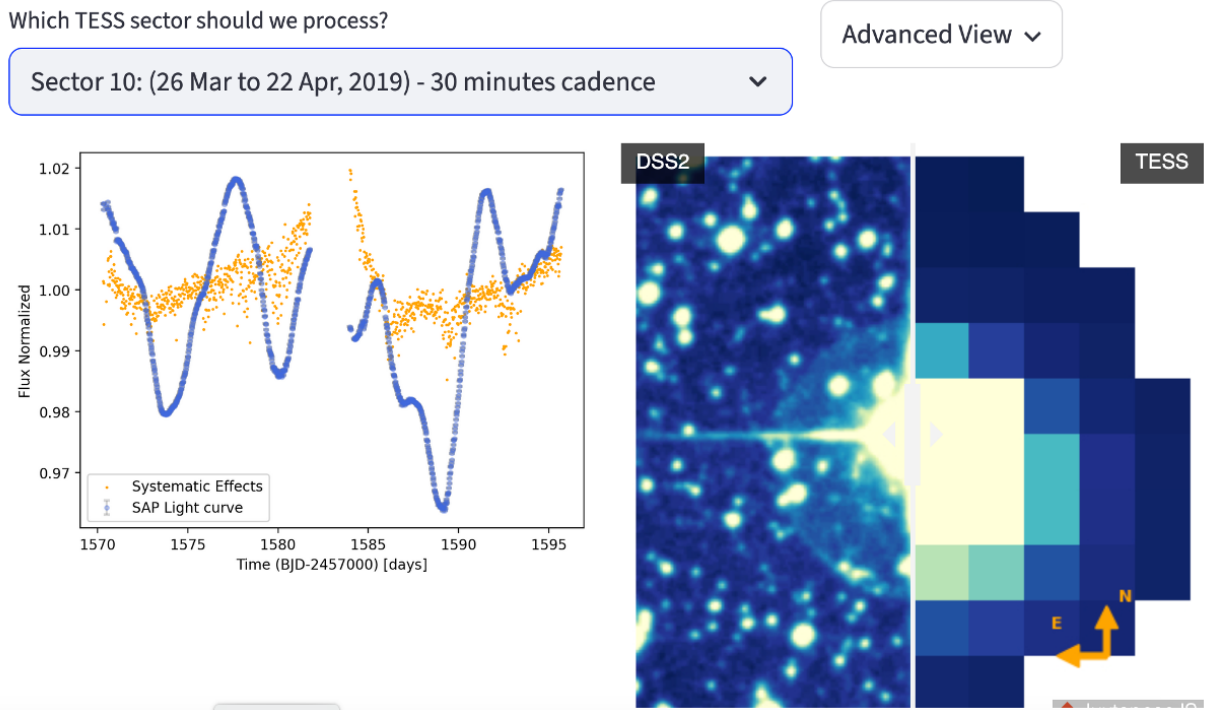

*Figure 4. Screen shot of TESS Extractor light curve for TIC 459537285, Sector 10, showing the simple aperture photometry light curve (blue), systematic effects used for detrending (yellow), and a split image of the Digitized Sky Survey-2 (DSS2)[7] field (left) and the TESS pixels sampled (right).*

We used TESS extractor to obtain at least one light curve for each of the 75 stars in our sample for initial screening. Our objectives were to maximize the possibilities for long-term ground-based monitoring, but also science return, so we were looking for:

- Bright stars (magnitude < 8)
- Larger-amplitude variations (greater than 0.02 mag)
- Quasi-periodicities longer than a day
- Evidence for changes in variability analogous to those seen in Deneb

---

[7] https://archive.eso.org/dss/dss/

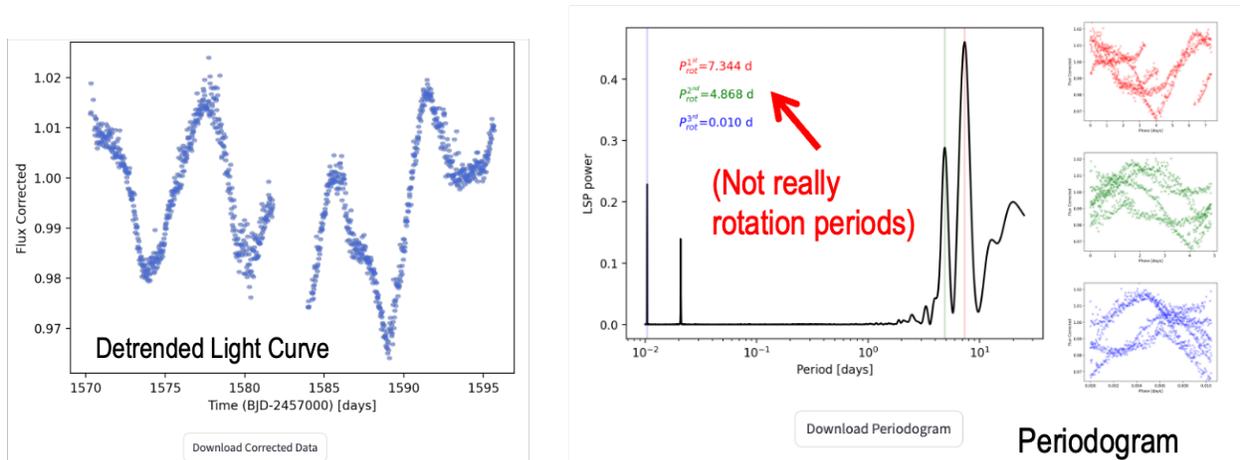

*Figure 5. Screen shot of TESS Extractor detrended normalized light curve for TIC 459537285, Sector 10, and Lomb-Scargle periodogram along with subplots phasing the data by period. This star is not strictly periodic, although two quasi-periods at 4.9 and 7.3 days are found in this 27-day light curve. The period at 0.010 days is possibly an alias of the 30-minute observing cadence.*

After previewing all 75 stars, we selected eight bright targets (TESS magnitudes 0.7–7.3) for further study. We also selected for follow-up two fainter (11[th] magnitude) stars near the south ecliptic pole having many consecutive sectors of TESS observations. Table 1 lists the ten targets selected. For these targets, we used TESS Extractor to process light curves for the available sectors as of August 2025 and downloaded the detrended light curve data. The light curve data files contain the time as barycentric Julian Date - 2457000, the normalized flux, and the flux uncertainty, typically around $10^{-4}$. Note that a variation of around 0.1 or 10% in flux corresponds to a variation of around 0.1 stellar magnitudes. Below, we show a few highlights from our light curve search. We chose to replot the light curves rather than use the TESS Extractor plots directly since we wanted to plot the data on the same x and y axis scale to better compare light curve variations between sectors.

*3.1 TIC 231308237 (Rigel)*

The well-known star Rigel in Orion, spectral type B81a, is an ACYG variable. It is the brightest of the ACYG variables, with TESS magnitude 0.706.

Figure 6 shows the normalized flux vs. time from TESS Extractor for TESS Sectors 5 and 32. The observing cadence is not the same for each Sector, but it appears that Rigel had entered a time of lower-amplitude and longer-period variability in Sector 32 compared to Sector 5. Without intervening observations, it is impossible to know whether and when Rigel's variability changed its character. We are hoping that Rigel can be monitored by ground-based observers, e.g., the AAVSO PEP team (see Calderwood and van Ballegoij, these proceedings).

We should also caution that we have not ruled out the possibility that the normalized flux from the TESS light curves for Rigel, or any other ACYG variable discussed here, is not actually varying between observed TESS sectors; the star's position on different CCD cameras and pixels in each sector, or changes in camera performance and background noise levels, could also change the normalized flux. Long-term ground-based monitoring and comparison of TESS Extractor results with those of other processing pipelines would help to confirm amplitude changes.

It is also likely that the default choice of TESS pixels to integrate over is not optimum for a star as bright as Rigel. The TESS Extractor and other software, e.g., lightkurve[8] do have the option to adjust the selection of pixels included in the analysis.

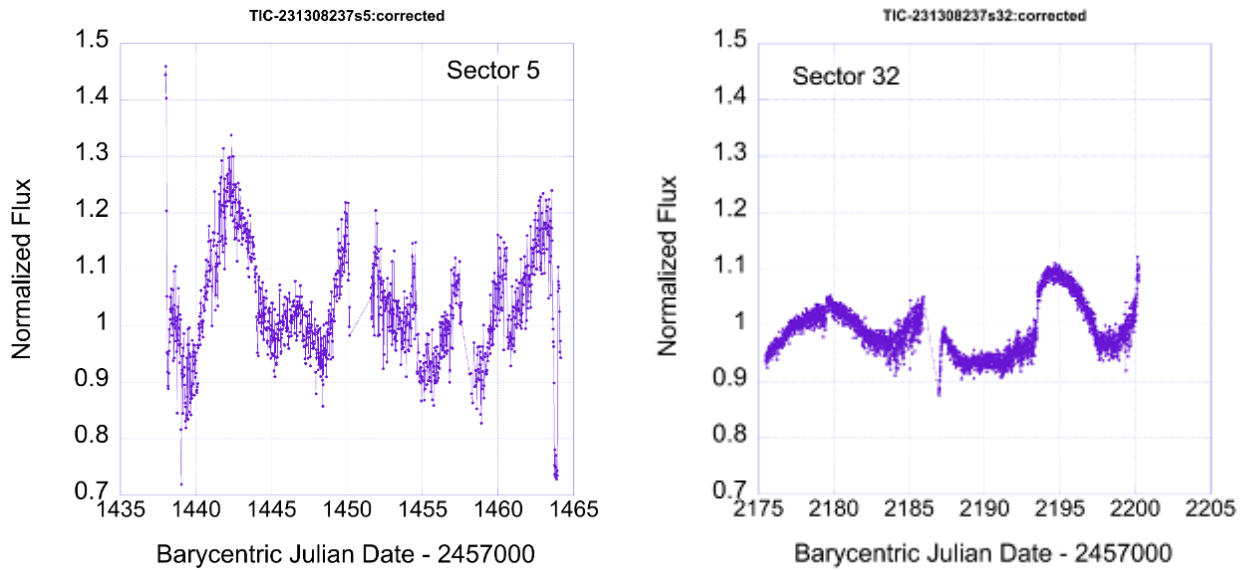

*Figure 6. Normalized flux vs. time for TESS Extractor data for TIC 231308237 (Rigel) for Sectors 5 and 32. The light curve appears more regular and seems to have decreased in amplitude in Sector 32. However, given the gap in the time series between Sector 5 and 32, it is not possible to determine when the change in light curve behavior occurred. See text for other caveats.*

*3.2 TIC 304894154*

This star has spectral type B3Ib and TESS magnitude 6.56. Figure 7 shows the TESS Extractor light curves on the same x and y scale for Sectors 10, 11, 63, 64, and 90. The variability appears to have lower amplitude in Sector 10, then increases in amplitude in Sectors 11, 63 and 64. There is a large excursion in Sector 90, followed by variability with a more defined period of around 3 to 4 days. Again, continuous coverage between these data gaps would have been useful to discern trends.

---

[8] https://lightkurve.github.io/lightkurve/

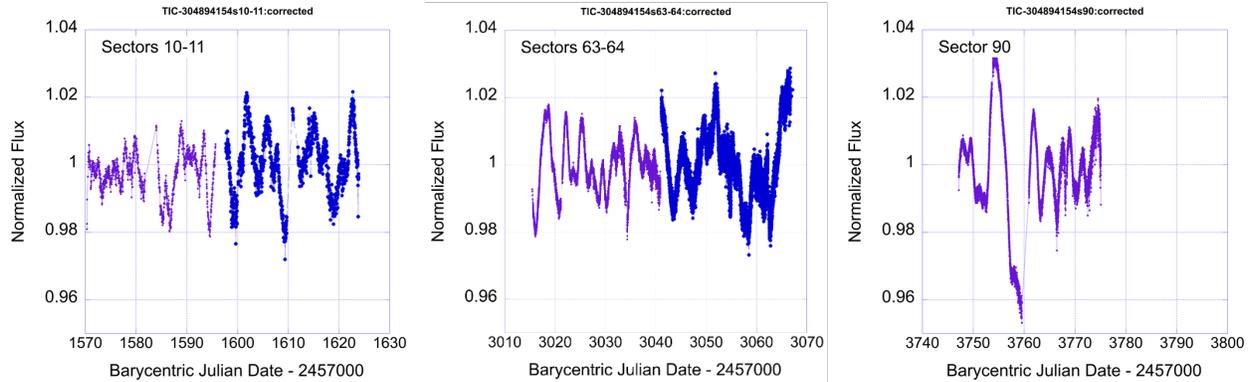

*Figure 7. Normalized flux vs. time for TIC 304894154 TESS Extractor data for Sectors 10, 11, 63 64, and 90. The light curve seems to have increased in amplitude during Sectors 11, 63, and 64, and there is a large excursion during Sector 90. The light curve is more regular/periodic during Sector 63 and the end of Sector 90.*

*3.3 TIC 457766442*

This star has TESS magnitude 5.61, and spectral type A2Iab, nearly the same spectral type as Deneb. Figure 8 shows the TESS Extractor light curves plotted on the same x and y axes scales for Sectors 36, 37, 63, 64, and 90. The amplitudes seem to be larger in Sectors 63-64. The Sector 63 data appear to be noisy, and we have traced this problem to noise in the TESS Extractor detrending algorithm for this sector—the noise is not present in the raw light curve or in light curves found on MAST using other processing pipelines. When the light curve is more regular, e.g., in Sectors 37 and 64, a periodicity of around 10 days can be discerned, similar to Deneb's 12-day quasi-period, which also varies in coherence with time. Again, without more continuous coverage, it is impossible to tell whether/when the amplitudes and periodicities changed character between sectors.

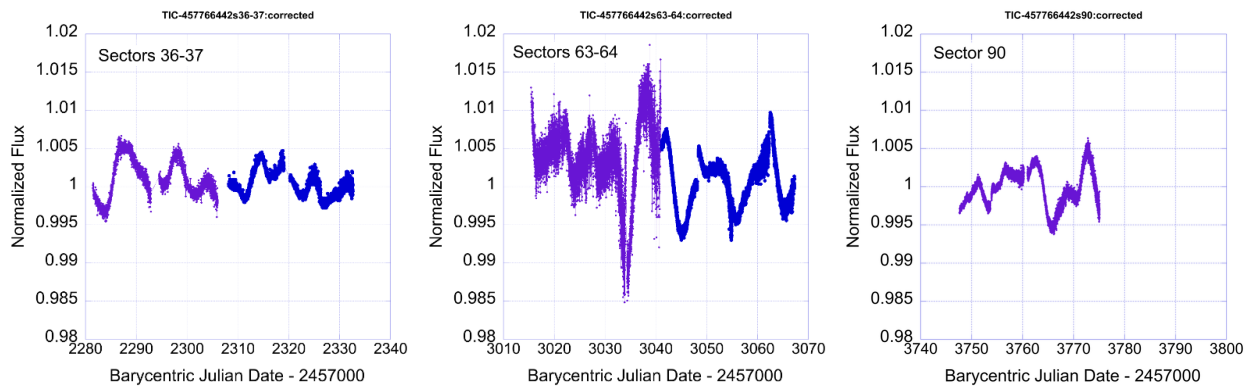

*Figure 8. Normalized flux vs. time for TESS Extractor data for TIC 457766442 for Sectors 37, 67, 63, 64, and 90. The light curve seems to have increased in amplitude in Sectors 63 and 64 compared to earlier and later sectors.*

## 4. ACYG variables on the Hertzsprung-Russell Diagram

We attempted to place the ten ACYG variables of Table 1 on the Hertzsprung-Russell (H-R) diagram relative to other named variable star classes. Figure 9 shows the variable star H-R diagram reproduced from Jeffery et al. (2025), with the location of the ten ACYG stars shown by red diamonds. We also included two estimates for the location of Deneb, reflecting the large uncertainty in Deneb's distance. These locations are only approximate, because it is not straightforward to calculate the star's luminosity. To do so requires knowing the star's distance, sometimes derived from parallax measurements, which have large uncertainties. Also, one needs to know the star's visual magnitude and apply a bolometric correction appropriate for the star's spectral type to obtain the absolute bolometric magnitude. We used the SIMBAD database, but also the MAST interface where the Gaia DR3 (Gaia collaboration 2023) and TESS Input Catalogs (Stassun et al. 2019) could be queried to find distance and effective temperature estimates. For most of the stars, there were several values in the literature, catalogs, or databases, which sometimes differed significantly from each other, so we used our judgement to choose a reasonable value. We estimated the Bolometric Correction (BC) using a plot of BC vs. spectral type found in Wikipedia[9] from the Landolt-Börnstein handbook (1982). We did not take into account interstellar reddening/extinction.

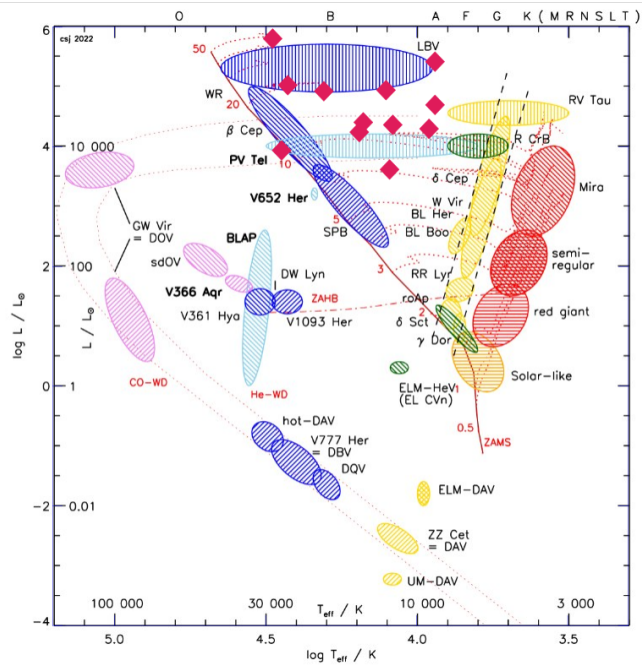

*Figure 9. Hertzsprung-Russell Diagram of pulsating variable star types reproduced from Jeffery (2025), Figure 1. Overlaid are estimated locations of ten ACYG stars from our study (red diamonds), along with two locations for Deneb based on two different distance estimates. The ACYG variables mostly are located below the LBV variables, to the right of the $\beta$ Cep variables, and left of the RV Tauri stars, but are close to the locations of these stars, indicating that the ACYG variables may be related to these variable star types in terms of variability mechanisms and evolutionary state.*

---

[9] https://en.wikipedia.org/w/index.php?title=Bolometric_correction&oldid=1317225345

ACYG variables may not be a homogenous class and may have different evolution states and causes of variability. Based on the spectral types and luminosities of well-known ACYG variables Deneb and Rigel on the H-R diagram, we expected to find ACYG stars below the Luminous Blue Variable (LBV) region, and between the β Cep and RV Tauri variable star regions. For the most part, the ten stars fell within this region. LBVs are supergiant stars more massive than ACYG variables which show microvariations, outbursts, and sometimes giant eruptions (Humphreys and Davidson 1994; Guzik and Lovekin 2014). RV Tauri variables are semiregular yellow supergiants which have light curves with alternating shallow and deep minima and periodicities of 20-90 days (Bódi and Kiss 2019). β Cep variables are main-sequence spectral type O and B stars of with masses 7-20 solar masses, which pulsate in multiple low-order radial and nonradial pressure and gravity modes, with periods of 0.1-0.3 days (Stankov and Handler 2005). The proximity of these variable star types on the H-R diagram to the ACYG variables may indicate some commonality in terms of mechanisms for variability.

**5. MAST High-Level Science Products**

Processed TESS light curves for many prior sectors for all 75 ACYG stars proposed for TESS Cycle 8 observations are available as HLSPs at the MAST portal.[10] Most light curve products are available as *fits* binary files, and a learning curve is required to locate, download and process the data using additional software tools. We developed a Jupyter Notebook[11] using astroquery[12] python tools to read the fits files and plot the light curves. Figure 10 shows a screen shot of the MAST web site. One can enter the target name, e.g., by TIC number, and view the list of data products available for download. We recommend selecting either TESS light curve products or target pixel files, and avoiding the full-frame image files, which have very large file sizes. There is also a 'cart' where files can be selected and placed for later bulk downloads.

*Figure 10. Screen shot of interface for TESS data search using MAST.*

---

[10] https://mast.stsci.edu/portal/Mashup/Clients/Mast/Portal.html
[11] https://jupyter.org/
[12] https://astroquery.readthedocs.io/en/latest/

The MAST offerings include many High-Level Science Products (HLSPs) created using different processing 'pipelines', which are not the same as the TESS Extractor pipeline. A non-exhaustive list of these includes:

- QLP: Quick Look Pipeline
- SAP/PDCSAP: Simple Aperture Photometry / Pre-Search Data Conditioning Simple Aperture Photometry
- SPOC: Science Processing Operations Center pipeline
- Eleanor Lite
- TASOC: TESS Asteroseismic Science Operations Center pipeline
- TGLC: TESS Gaia Light Curve, makes use of Gaia data for crowded fields

Except for the Quick Look Pipeline data, light curves are available for only a few stars for a few sectors. Table 1 lists available sectors for our ten stars from the QLP, PDCSAP, and Eleanor Lite pipelines. SPOC data were available for only TIC 30106140, and TASOC and TGLC data were available only for the faintest two stars. *An advantage of proposing stars to observe via the TESS General Investigator Program is that the SAP/PDCSAP-processed light curves are made available on MAST a few months after the observations are taken.*

Some pipelines are preferable to others for studying stellar variability. To intercompare results for different pipelines, for our ten stars of interest (Table 1) we examined light curves for the same sector using several pipelines. Note that even the 'raw' light curves shown below usually have been pre-processed in some way, e.g., normalized, artifacts and trends common to most pixels in the field removed, etc.

Figure 11 compares the 'raw' and detrended light curve for the Quick Look Pipeline for Sector 68 for TIC 179508685. The raw light curve (left) shows that the star has a 1.658-day periodicity, along with a longer higher-amplitude modulation superimposed. However, the Quick Look Pipeline detrending algorithm is intended to remove stellar variability to facilitate searching for planet transits, so one can see that the detrended light curve (right) is nearly flat.

Figure 12 shows the raw and detrended light curves for TIC 179508685 for Sector 10 from the Eleanor Lite pipeline. While the raw light curve has some artifacts at the beginning of the first and second half of the sector, the 1.658-day period is visible as well as the longer-period modulation. The Eleanor Lite corrected light curve has retained the shorter periodicity but has removed the longer-period modulation. The Eleanor Lite pipeline also appears to have introduced some noise and did not remove anomalous data at the beginning of the first and second half of the observing sector.

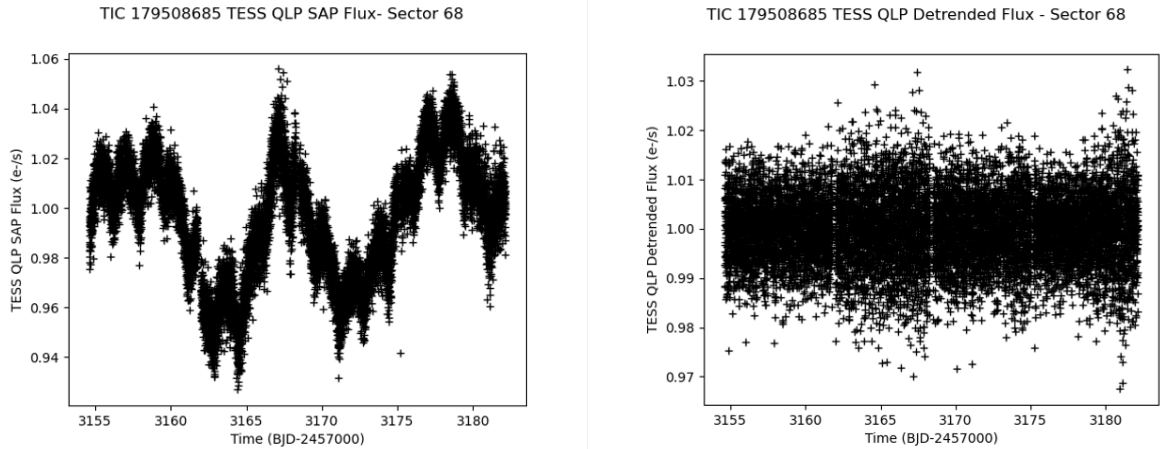

*Figure 11. Raw (left) and detrended (right) TIC 179508685 light curves for Sector 68 from the Quick Look Pipeline. A 1.658-day periodicity along with a longer-period modulation can be seen in the raw data; however, the detrended data removes both the short and long-period variability.*

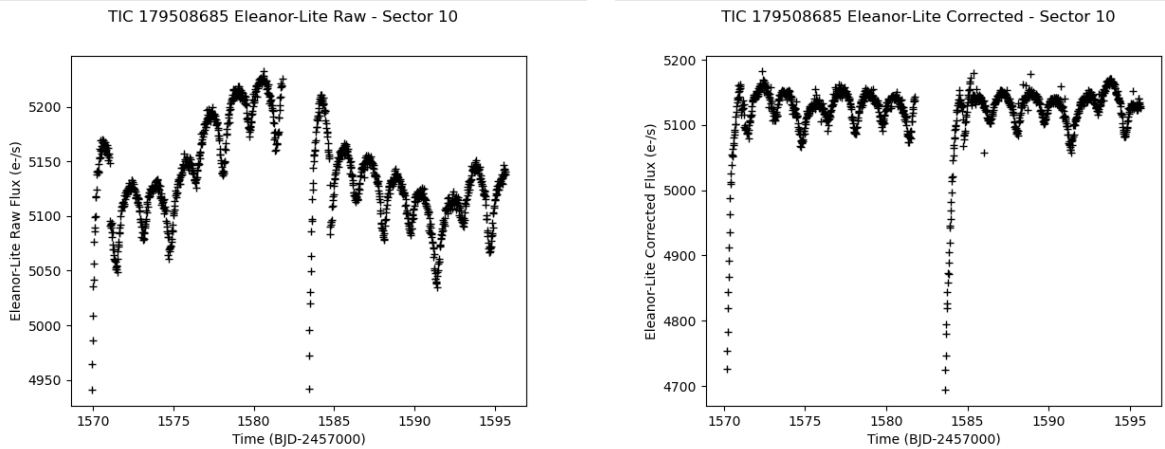

*Figure 12. Raw (left) and corrected (right) light curves from Eleanor Lite pipeline for TIC 179508685 for Sector 10. A 1.658-day periodicity with a longer-period modulation can be seen in the raw data. The Eleanor Lite corrected light curve flattens this long-term trend while retaining the shorter-period variation. The corrected light curve also retained artifacts at the beginning of each half-sector and has a few data points that must be cleaned.*

We next discuss the SAP/PDCSAP pipeline light curves that are made available for stars targeted as part of the TESS General Investigator program. Figure 13 shows the SAP (raw) and PDCSAP (corrected) light curve for TIC 459537285 during Sector 10. The light curve is not normalized, but it appears that the flux levels are shifted in the PDCSAP data. There seems to have been a slight amount of detrending, and small light curve segments in the PDCSAP data have been deleted at the beginning of each half-sector, even though these data do not look anomalous in the SAP light curve. While the PDCSAP light curve is usable and does not seem to have smoothed out real variations, we recommend comparing the SAP and PDCSAP light curves; our prior experience with SAP/PDCSAP products for Cepheid variable light curves showed that the PDCSAP pipeline sometimes removed real periodic variations.

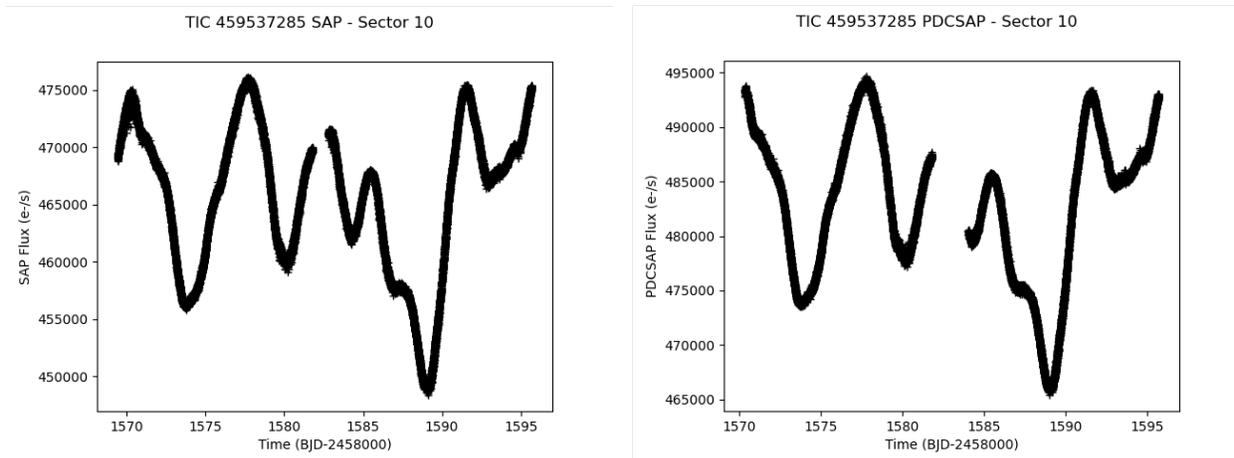

*Figure 13. TIC 457537285 Sector 10 SAP (left) and PDCSAP (right) light curves. The two light curves are very similar, except that the PDCSAP flux is higher and segments of data at the beginning of each half sector have been deleted.*

Figure 14 shows the 'raw' and 'corrected' light curves from the Eleanor Lite pipeline for TIC 459537285 during Sector 10. The raw Eleanor Lite light curve seems to have an obvious upward trend. The corrected light curve was so noisy that we had to replot the data and zoom in manually. The trend is flattened in the corrected data, but more cleaning would be needed to remove bad data points. Note that Figure 14 and Figure 13 can be compared to see the results for this star using two different pipelines. While some of the same features can be discerned, there are also significant differences, emphasizing the need for accurate long-term monitoring of these stars to establish an absolute amplitude scale.

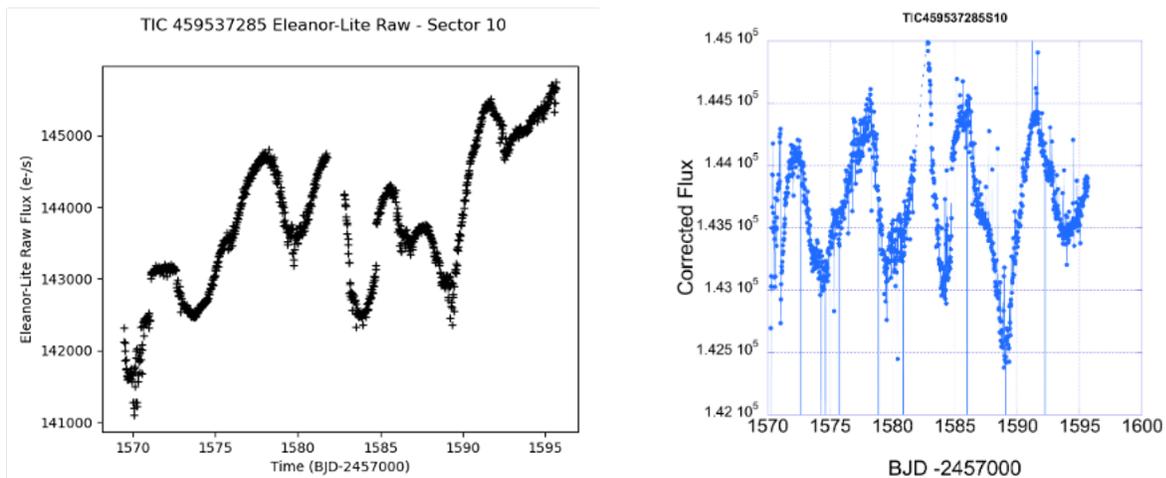

*Figure 14. TIC 4959537285 Sector 10 Eleanor-Lite pipeline raw (left) and corrected (right) light curves. The raw light curve has an upward trend that was removed in the corrected light curve, but the corrected curve has many bad data points that need to be cleaned. The corrected light curve is very similar to the SAP and PDCSAP light curves of Figure 13.*

To summarize our recommendations regarding light curves processed using various data pipelines available on MAST:

***Quick Look Pipeline (QLP):*** The raw light curve is useful for a 'quick look' but may still contain trends. However, the detrended light curve is not useful for variable stars since the long-term trends and pulsation-like variations are removed to optimize searches for transiting exoplanets.

***SAP/PDCSAP:*** This pipeline is very useful, and light curves are made available on MAST for all targets requested via the General Investigator program. We recommend comparing the SAP (simple aperture photometry) and PDCSAP (corrected) light curves to be sure that real trends and pulsations have not been removed in the PDCSAP light curve.

***Eleanor Lite:*** This pipeline removes longer-period trends which may be real from the corrected light curve but retains shorter-period variations. Also, the resulting corrected light curves are often noisy and have bad data points, requiring further cleaning.

*Table 1. Ten ACYG variables selected for follow-up after screening light curves using TESS Extractor. TESS Magnitudes are from the TESS Input Catalog 8.2 (Stassun et al. 2019). Spectral types are from the SIMBAD database.[13] High-Level Science Product light curves processed by several different pipelines are available for some of the stars and some sectors at MAST.[14]*

| TIC | TESS Magnitude | Spectral Type | TESS Extractor | QLP | SAP/PDC SAP | Eleanor Lite | Notes |
|---|---|---|---|---|---|---|---|
| | | | Sectors Available | | | | |
| 231308237 | 0.706 | B8Ia | 5, 32 | 5, 32 | | 5 | Rigel |
| 317046099 | 6.57* | B2Ia | 10, 11, 37, 38, 64 | 10, 11, 37, 38 | 10, 37, 38, 64 | 10 | V808 Cen |
| 337793038 | 5.24 | O8Iaf | 12, 39, 66, 90 | 12, 39, 66 | 12, 39, 66 | | V973 Sco |
| 457766442 | 5.61 | A2Iab/b | 36, 37, 63, 64, 90 | 36, 37, 63, 64 | 36, 37, 63, 64 | | |
| 459537285 | 8.66* | B8Ia | 10, 36, 37, 63, 64, 90 | 10, 36, 37, 63, 64 | 10 | 10 | V523 Car |
| 304894154 | 6.56 | B3Ib/II | 10, 11, 63, 64, 90 | 10,11,63,64 | 10 | 10 | |
| 458076250 | 6.64 | B0.7Ib | 10, 36, 37, 63, 64, 90 | 36, 37, 63, 64 | | | V513 Car |
| 460279488 | 7.32 | B8Iab | 10, 11, 36, 37, 63, 64, 90 | 10, 11, 36, 37, 63, 64 | | 10 | V508 Car |
| 30106140 | 11.037 | B5Ia+ | 1-39, 61-69, 87-90 | 32 sectors, 1-69 | 23 sectors, 27-88 | 1,2,3,4, 6,7,8,9, 10 | |
| 179508685 | 11.365 | OB D | 1-13, 27-39, 61-69, 87-90 | 30 sectors, 2-69 | | 2,3,4,5, 6,7,8,9, 10 | 1.658-day period with longer-period modulation |

*TESS Input Catalog magnitudes were not correct; V magnitude from SIMBAD is listed instead.

We did not show example light curves from the SPOC, TASOC, TGLC pipelines here. The SPOC pipeline is the overarching processing pipeline that produces the SAP/PDCSAP light curves, and it is not clear what the differences might be between SPOC and SAP/PDCSAP light-curve products. Of the ten stars in Table 1, only TIC 30106140 had SPOC light curves available on MAST. The TASOC pipeline is excellent for pulsating variable stars, but light curves are not available on MAST for most ACYG stars and are available only for early TESS sectors. The TGLC pipeline makes use of Gaia DR3 (Gaia Collaboration 2023) data for position corrections for stars in crowded fields and clusters and looks promising. For the ten stars in Table 1, TGLC light curves are available only for the faintest two stars and for early sectors.

---

[13] https://simbad.u-strasbg.fr/simbad/
[14] *https://mast.stsci.edu/portal/Mashup/Clients/Mast/Portal.html*

## 6. Summary and Future Work

We are only in the initial stages of exploring the large number of TESS light curves available for ACYG variables. However, the TESS time-series length of 27 days per sector leaves long gaps between observations for most of the targets. Detrending the light curves and placing them on the same absolute scale to compare sectors is challenging. Longer-term, more continuous monitoring by ground-based observers as well as observations overlapping in time with the TESS observations are needed to fill in the time-series gaps and verify changes in amplitude or large excursions found in the TESS data.

We plan to continue analysis of the existing TESS light curves for all 75 targets proposed for TESS Cycle 8, including the new light curves from Cycle 8 as they become available. Even though ACYG variables are not strictly periodic, we plan to examine Fourier transforms of existing time-series data to determine whether significant quasi-periodicities are present. We also plan to identify ACYG stars to propose for TESS General Investigator Cycle 9 observations.

We have identified a few ACYG stars for follow-up by the AAVSO ground-based PEP observing program. We also plan to search for ACYG light curve data from other sources, such as from the prior NASA *Kepler* (Borucki et al. 2008) and K2 (Howell et al. 2018) missions, and the Solar Mass Ejection Imager (Jackson et al. 2004, Clover et al. 2011, Guzik et al. 2025).

We also plan stellar evolution and pulsation modeling (see Moore and Guzik 2025). We hope that these studies will lead to a better understanding of the origin of the variability and the evolutionary state of Deneb and the ACYG variables.

## Acknowledgements

This work makes use of data from NASA's Transiting Exoplanet Survey Satellite (TESS), obtained from the Mikulski Archive for Space Telescopes (MAST). We also acknowledge use of Gaia DR3 and SIMBAD databases. We made use of open-source software including astroquery python tools and TESS Extractor. We thank Philip Masding (British Astronomical Association) for suggesting TESS Extractor, and Javier Serna for answering our queries. We acknowledge support from Los Alamos National Laboratory, managed by Triad National Security, LLC, for the U.S. DOE's NNSA Contract #89233218CNA000001. This research has made use of the International Variable Star Index (VSX) database, operated at AAVSO, Cambridge, Massachusetts, USA.